# Bipolar and unipolar electrical fatigue in ferroelectric lead zirconate titanate thin films: an experimental comparison study

X.J. Lou[*] and J. Wang

*Department of Materials Science and Engineering, National University of Singapore, 117574, Singapore*

**Abstract:**

By performing standard PUND (positive-up-negative-down), hysteresis-loop and dielectric measurements on the ferroelectric lead zirconate titanate (PZT) thin-film capacitors subject to bipolar/unipolar electrical cycling, we show that unipolar fatigue is evident though still less severe than bipolar fatigue conducted at the same voltage. That has been attributed to polarization retention (backswitching) induced by the residual depolarization field between the monopolar pulses where the applied field is lower than the depolarization field, and explained using the LPD-SICI model (LPD-SICI stands for local phase decomposition caused by switching-induced charge injection). The conventional view that switching does not occur during unipolar electrical cycling may need to be corrected. PUND results recorded using the pulses of the same voltage as those for repetitive fatigue cycling are not reliable if the voltage is lower than $2V_c$ ($V_c$ is the saturated coercive voltage). Dielectric measurements or hysteresis-loop measurements at higher voltages (e.g. $4V_c$) are more reliable ways to evaluate the degree of fatigue and could provide more valuable information in such situations. Finally, dielectric results have been used to estimate the *effective* thickness $d_i$ of the fatigue-induced degraded (pyrochlorelike) interfacial layer after bipolar/unipolar fatigue, which has not been done so far to our best knowledge. The fact that $d_i$ is still much less than the film thickness even after the most severe bipolar fatigue strongly suggests that polarization fatigue in ferroelectrics is an interface effect, not a bulk one.

---

[*] Electronic mail: mselx@nus.edu.sg

## I. Introduction

Polarization fatigue in ferroelectric materials is defined as a systematic suppression of spontaneous polarization under bipolar or unipolar electrical cycling.[1, 2] Many publications in the past deal with bipolar electrical fatigue in ferroelectric thin films,[2] and fewer deal with unipolar fatigue in both thin-film and bulk ceramic ferroelectrics.[3-5] This is probably because of the potential applications of ferroelectric memories (FeRAM) in future non-volatile memory devices, where each cell is driven by bipolar pulses. However, it should be noted that in case of the devices made from ceramic ferroelectrics the driving pulses used are usually unipolar. So a good understanding of unipolar fatigue is critical in solving the reliability issues in these devices. On the other hand, careful experimental studies of unipolar fatigue in ferroelectrics may shed new light on this phenomenon from a different perspective, helping us tell which fatigue mechanism(s) really sits behind.

Although it has been generally reported that unipolar electrical cycling results in no or much less pronounced fatigue in ferroelectric materials in comparison with bipolar cycling,[3-5] a systematic comparison study of both bipolar and unipolar fatigue in thin-film ferroelectric capacitors subject to a range of driving fields (e.g. $E_c$, $2E_c$, $3E_c$ and $4E_c$; $E_c$ stands for the *saturated* coercive field) has not been done. In particular, early works usually evaluated the degree of unipolar/bipolar fatigue of a specific sample by PUND measurements (PUND denotes the Positive-Up-Negative-Down five-pulse measurement termed by Radiant Technologies; this method will be explained further later on)[6] or direct hysteresis-loop measurements[2] after a certain number of switching cycles. In this paper, we show that neither the PUND measurement nor the hysteresis measurement alone can give all the information about the fatigued state of a ferroelectric capacitor after bipolar or unipolar electrical cycling, and in some cases they even give contradictory conclusions as we will show later on. To get a full picture out of it, dielectric measurements employed by other researchers[7-12] in the past must be conducted. Finally, we show that the dielectric data recorded from the virgin and fatigued states of a ferroelectric capacitor could be used to estimate the effective thicknesses of the degraded interfacial layers after a certain number of electrical cycling, which could not be done based on our earlier micro-Raman results on the fatigued PZT films.[13]

## II. Experiment

The $Pb(Zr_{0.3}/Ti_{0.7})O_3$ (PZT) thin films of ~300 nm in thicknesses used in the present work were fabricated on $Pt/TiO_x/SiO_2/Si$ substrates using sol-gel spin coating method. 10% Pb-excess solutions of 0.4 M was used to prepare the PZT films in order to compensate Pb loss during annealing. After a desired number of layers were coated, the PZT films were finally annealed at 650 ºC for 15 min in air in a quartz tube furnace (Carbolyte) to achieve the wanted phase. The preparation procedures are as the same as those reported elsewhere.[14] X-Ray Diffraction (XRD, D8 Advanced Diffractometer System, Bruker) and Field Emission Scanning Electron Microscopy (FE-SEM, XL30 FEG Philips) studies show that the films are polycrystalline and of a perovskite phase. For electrical measurements, the top electrodes (Pt squares of ~100 x 100 $\mu m^2$) were then deposited by sputtering on the films via transmission electron microscopy (TEM) grids. Standard polarization fatigue (i.e., the cycling-and-PUND procedure) and hysteresis-loop measurements were carried out using a Radiant ferroelectric test system. Dielectric measurements were performed using an impedance analyzer (Solartron SI 1260).

## III. Results and discussion

### A. Bipolar fatigue

To help those who are not quite familiar with fatigue measurements understand better the context of this paper as well as how the standard fatigue measurement is carried out, a brief introduction of the cycling-and-PUND method developed by Radiant Technologies may be necessary at this stage.

To measure the fatigue properties of a sample, one has to develop a method to extract its remanent polarization. In the Radiant RT6000 system, a PUND (Positive-Up-Negative-Down) measurement is employed to carry out this task. A PUND measurement is a standard pulse measurement consisting of five pulses of identical pulse magnitude and width (see Fig 1).[6] The first pulse presets the polarization state of a ferroelectric sample to a given direction. The second pulse switches the sample to the opposite state and the switched polarization value is extracted. A second non-switching measurement is performed by applying the third pulse identical to the second one after it returns to zero volts and a programmable delay period passes. The delay is intended to allow all the non-remanent polarization components to settle before the next measurement starts. In the terminology used by the Radiant ferroelectric system for PUND measurements, $P^*$ and $P^*_r$ represent the switchable polarization containing both remanent and non-remanent polarization, while $P^\wedge$ and $P^\wedge_r$ represent the non-switchable polarization containing only non-remanent polarization components.[6] $dP = P^* - P^\wedge$ (very close to $dP_r = P^*_r - P^\wedge_r$), denoting the non-volatile polarization $P_{nv}$ or the *truly* switchable polarization $P_{sw}$ (~$2P_r$; $P_r$ is the remanent polarization extracted from hysteresis-loop measurements). The fourth and fifth pulses mirror the second and third pulses, but point in the opposite direction. These measurements give $-P^*$, $-P^*_r$, $-P^\wedge$ and $-P^\wedge_r$. Therefore, a fatigue measurement simply becomes a series of stress/measure events consisting of a period of application of the fatigue stress with a certain number of electrical cycles, followed by PUND measurements. To determine whether fatigue has occurred or not, it is necessary to carry out an additional PUND measurement on the virgin sample to provide a pre-fatigue (virgin) baseline, against which the results obtained from the bipolar/unipolar stressed sample are compared.

Fatigue process is generally characterized in the literature by displaying its growth (e.g. $dP$, $P^*$, $P^\wedge$, or $P_r$) with the total number of bipolar/unipolar waveform cycles in a logarithmic plot.[2] This suggests that some adjustment has to be made, i.e., each subsequent measurement would be of longer duration than the previous one. So the number of cycles between two PUND measurements greatly increases as the fatigue process continues. The whole algorithm for bipolar fatigue measurements is shown in Fig 1 (The one for unipolar fatigue is shown in Fig 5). The waveform depicted in Fig 1 (and Fig 5) and chosen for fatiguing our samples in the present work is triangular wave, which can be readily defined by specifying the voltage and frequency. Other options include sine wave, square wave, pulse train and arbitrary custom waveform.[6]

In general, the voltage of the PUND pulses is set to be the same as that of the cycling wave. However, we will show in this paper that in some situations this methodology might be problematic and not be able to give us the correct information about the genuine state of a fatigued ferroelectric capacitor. To show this problem more clearly, additional hysteresis-loop and dielectric measurements on the progressively fatigued PZT thin film were carried out. The way of resolving this issue in the cycling-and-PUND procedure will be discussed and the methodology of revealing the real state of a fatigued ferroelectric memory cell will be provided.

Bipolar fatigue profiles of four Pt/PZT/Pt thin-film capacitors on the same film evaluated by PUND measurements are shown in Fig 2. Fig 2(a) shows $P_{sw}(\equiv dP$, extracted from PUND measurements) as a function of cycle number $N$ for bipolar fatigue at 3 V (equivalent to $E_{appl}$~100 kV/cm~$E_c$; see Fig 3), 6 V ($E_{appl}$~200 kV/cm~$2E_c$), 9 V ($E_{appl}$~300 kV/cm~$3E_c$) and 12 V ($E_{appl}$~400 kV/cm~$4E_c$) driven by triangular wave of $10^5$ Hz in frequency; Fig 2(b) displays the normalized results for the data shown in Fig 2(a) (Note that the voltage of the PUND pulses was set to be *the same* as that of cycling wave as shown in Fig 1 and mentioned above). One can see that the fatigue characteristics of our PZT film show a logarithmic dependence on cycle number and polarization fatigue becomes more severe and eventually saturated for the capacitors driven by higher cycling voltages, consistent with the general features reported in the literature.[2, 15-19]

The electric field/voltage dependence of fatigue data in Fig 2 can be well explained by the LPD-SICI model (LPD-SICI stands for local phase decomposition initiated by switching-induced charge injection):[2, 20] the higher fatigue rates at higher applied voltages are caused by higher $P_r(E_{app})$ and consequently more intensive charge injection and phase decomposition at the domain nucleation sites at the electrode-film interface.[20] The saturation (or overlapping) of the fatigue profiles for 6, 9 and 12 V can be explained by the saturation of $P_r(E_{app})$ for $E_{app} \geq 2E_c$ using this model.[20]

The only thing that looks a little unusual in Fig 2 is the fatigue profile for 3 V, in which more than 90% of $P_{sw}(0)$ is lost after $10^9$ cycles. This seems too significant, given that 3 V is just about the saturated coercive voltage for the film under study. Under a high cycling frequency of $10^5$ Hz (i.e., ~10 ms for each cycle) the capacitor is supposed not to be fully switched at such a low applied field (~$E_c$).[15, 21, 22] Therefore the real conditions of the 3 V-fatigued sample are probably much better than that shown in Fig 2.

Fig 3 show the fatigue characteristics of these four capacitors evaluated simultaneously by hysteresis-loop measurements using triangular wave of 1000 Hz in frequency and 12 V in magnitude (i.e., ~$4E_c$) after 1, $4 \times 10^5$, $1.9 \times 10^6$, $1.15 \times 10^7$, $1.02 \times 10^8$ and $1 \times 10^9$ cycles for bipolar fatigue at 3 V [Fig 3(a)], 6 V [Fig 3(b)], 9 V [Fig 3(c)] and 12 V [Fig 3(d)]; Fig 3(e) illustrates the normalized results for $2P_r(N)$ as a function of cycle number $N$; $2P_r(N)$ is extracted from the loops shown in Fig 3(a-d). Indeed, we see that although the driving voltages of 6, 9, and 12 V gave rise to significant fatigue in the film [i.e., ~90% of $2P_r(0)$ was lost after $10^9$ cycles; the higher the cycling voltage the poorer the fatigue endurance will be] in good agreement with the data shown in Fig 2(a)(b), 3 V cycling voltage only resulted in polarization loss of ~35% after $10^9$ cycles [see Fig 3(a) and Fig 3(e)], in complete disagreement with the 3 V bipolar fatigue data shown in Fig 2 (a)(b). This justifies the conventional view in the literature that the driving electric field employed for fatigue measurements should be $\geq 2 E_c$. (equivalent to $\geq 6$ V in this work).

Note that the minor difference between the normalized fatigue profiles at 6 V (or 9 V, 12 V) shown in Fig 2(b) and Fig 3(e) is due to the difference between the methods used to extract the residual polarization of the *same* capacitor undergoing bipolar fatigue: PUND measurements were employed to get $P_{sw}(N)$ shown in Fig 2(b), while standard hysteresis-loop measurements were used to extract $2P_r(N)$ shown in Fig 3(e); $P_{sw}(N)$ is around but slightly less than $2P_r(N)$, because the former contains switchable polarization only as aforementioned while the latter still contains a small amount of non-switchable polarization.

To obtain more information from these four capacitors undergoing bipolar fatigue, their dielectric characteristics were also investigated and are shown in Fig 4 [(a)(b) for 3 V, (c)(d) for 6 V, (e)(f) for 9 V and (g)(h) for 12 V]. Dielectric measurements were carried out at a frequency range of $10^1$ to $10^6$ Hz with an ac signal of 0.05 V in amplitude. Fig 4(a)(c)(e)(g) show the data for dielectric constant and tan loss as a function of measurement frequency; and Fig 4(b)(d)(f)(h) display the results for dielectric constant and tan loss measured at $10^2$ and $10^5$ Hz, respectively, as a function of cycle number $N$. Fig 4(i) shows the normalized dielectric constant measured at $10^5$ Hz as a function of cycle number $N$. Again, we see that dielectric constant decreases as a function of switching number either slightly or dramatically depending on the magnitude of driving voltage [see Fig 4(a-i)]; the higher the cycling voltage the more considerable the decrease in dielectric constant will be and the most significant drop in dielectric constant occurs for the capacitor driven by 12 V, e.g., ~50% of the overall dielectric constant of the virgin capacitor was lost after $10^9$ cycles [see Fig 4 (i)].

The degradation in dielectric constant during bipolar fatigue observed in this work is consistent with those reported by other researchers for both ferroelectric thin films[7-9] and bulk ferroelectrics.[10-12] This phenomenon was explained previously by the experimental results showing that the fatigue-induced phase-decomposed interfacial layer may be pyrochlorelike and has dielectric constant of ~30 only.[2, 13] We also noticed that although the tangent loss measured at $10^5$ Hz keeps almost constant the one measured at $10^2$ Hz increases progressively with the number of cycles, particularly for bipolar fatigue at 9 and 12 V [Fig 4(a-h)]. This phenomenon could be interpreted via the accumulation of space charges at the degraded interface caused by charge injection according to the LPD-SICI model.[20] These charges could only respond to the external stimulus when frequency is low enough (e.g., ~$10^2$ Hz).

We also noticed that dielectric constant of the capacitor undergoing bipolar fatigue at 3 V decreases only slightly (~13%) after $10^9$ cycles [Fig 4 (b)(i)], consistent with the hysteresis-loop measurements shown in Fig 3 (a)(e), but in total disagreement with the PUND results shown in Fig 2(a)(b). This, again, confirms our argument made above that fatigue voltage should be $\geq 2\ E_c$. For $E_{appl}$~

$E_c$ or $E_{appl} < E_c$, the common PUND setting-up (where the voltage of PUND pulses is set to be the same as the cycling voltage) is unable to give correct information about the real state of the fatigued capacitor. To solve this problem concerning low-voltage/field fatigue studies, we could make use of the following evaluation methods: (1) enhancing the voltage of PUND pulses. For the capacitors in this work, we could set the PUND pulse voltage to be 9 V or 12 V instead of 3 V for bipolar fatigue measurements conducted at 3 V cycling voltage; (2) conducting hysteresis-loop measurements at higher voltages (e.g., 9 V or 12 V) as we did in this work and shown in Fig 3 (a-e); (3) performing dielectric measurements to evaluate the change in the overall dielectric constant and tangent loss of the capacitor after progressive electrical fatigue [e.g., see Fig 4 (a-i)].

**B. Unipolar fatigue**

Fig 5 shows the schematic of a standard unipolar fatigue measurement employed by Radiant Test Systems. One can see that it follows almost the same algorithm as the standard bipolar fatigue measurements shown in Fig 1 does. Triangular wave of $10^5$ Hz was used for both bipolar and unipolar fatigue measurements in this work for good comparison. Note that the set-up of unipolar fatigue is a little different from that of bipolar fatigue. For unipolar fatigue carried out using triangular wave and the like, both the waveform magnitude and the waveform offset should be set to be half of the desired fatigue voltage.[23] For instance, a triangular wave with amplitude of 6V and offset of 6V will lead to a triangular wave centered at 6V with +/-6V amplitude. So we obtain a 12 V unipolar triangular wave changing periodically from 0 V to 12 V and then back to 0 V (see Fig 5; unipolar fatigue at 3, 6 and 9 V could be set in a similar manner).

Following exactly what we did for bipolar fatigue in Fig 2-4, we carried out unipolar fatigue on another four virgin capacitors on the same film at 3, 6, 9, 12 V, respectively. The unipolar fatigue characteristics of these thin-film capacitors at different cycling voltages are shown in Fig 6 (evaluated by standard PUND measurements), Fig 7 (evaluated by hysteresis-loop measurements at 12 V after 1, $4\times10^5$, $1.9\times10^6$, $1.15\times10^7$, $1.02\times10^8$ and $1\times10^9$ cycles) and Fig 8 (evaluated by dielectric measurements), respectively. Detailed descriptions of these curves are given in the captions of these figures.

The combination of Fig 6-8 shows that the unipolar fatigue resistance of these capacitors decreases as the fatigue voltage increases, displaying the same trend as the bipolar fatigue resistance does (see Fig 2-4); the saturation of the fatigue curves at higher voltages such as 9 and 12 V could also been seen from these figures. Again, the inconsistency in unipolar fatigue at 3 V evaluated by PUND measurements [most than 85% of $P_{sw}(0)$ was lost after $10^9$ cycles, Fig 6 (a)(b)] and by hysteresis-loop measurements [less than 4% of $P_r(0)$ was lost after $10^9$ cycles, Fig 7 (a)(e)] and by dielectric measurements [~7% of dielectric constant was lost after $10^9$ cycles, Fig 8 (a)(b)(i)] could also been clearly seen, as the same as the case for bipolar fatigue at 3 V discussed above. Note that here we see the similar agreement between hysteresis-loop measurements and dielectric measurements for 3 V unipolar fatigue as we saw for 3V bipolar fatigue shown above. We attribute the reason for no or little fatigue at 3 V to the low applied unipolar (or bipolar) field (3 V~$E_c$), which could not fully switch the Pt/PZT/Pt capacitor with inferior "dead layers" at the interface and thus large depolarization field $E_{dep}$ within the film under a high frequency of $10^5$ Hz.[15, 21, 22] This issue will be discussed further in the next section.

**C. Comparison of bipolar and unipolar fatigue in ferroelectric thin films: what factor matters?**

Having discussed the fatigue characteristics of the capacitors undergoing bipolar and unipolar fatigue at various voltages such as 3, 6, 9 and 12 V, we now turn to the difference between bipolar and unipolar fatigue in a ferroelectric thin film, mainly concentrating on the data recorded at the *same* voltage for comparison.

By comparing Fig 2 with Fig 6, Fig 3 with Fig 7 and Fig 4 with Fig 8, one can see that under the *same* cycling voltage unipolar fatigue is indeed less severe than bipolar fatigue, consistent with experimental results reported previously.[3-5] For instance, we see that the polarization loss is only less than 4% after $10^9$ unipolar cycles at 3 V [Fig 7 (a)(e)] while ~35% of $P_r(0)$ is lost after $10^9$ bipolar cycles at the same voltage [Fig 3(a)(e)]. Although the unipolar fatigue endurance is significantly reduced at higher voltages such as 6, 9 and 12 V, it still shows much better performance than the bipolar one at the same voltage. However, the noteworthy unipolar fatigue rate at relatively higher voltages looks a little "unusual" in comparison with the previous results[3-5] and deserves further explanations.

It is well known that polarization fatigue in ferroelectrics (or antiferroelectrics) is closely related to switching mechanism in ferroelectrics[2, 20] (or antiferroelectrics[24]). Therefore, to solve the former the latter has to be properly understood. Since the first work was done by Merz in 1950s,[25] polarization switching in ferroelectrics has been intensively studied both experimentally[26-29] and theoretically[21, 22, 30-33] for a few decades. Here we attribute the reason for the evident unipolar fatigue observed in this work on the Pt/PZT/Pt thin-film capacitors to a series of periodic events of polarization backswitching and switching subject to unipolar driving triangle wave. "Backswitching" during unipolar fatigue is defined as a polarization switching process driven by the residual depolarization field $[E_{dep}(t)-E_{appl}(t)]$ when the applied unipolar field $E_{appl}(t)$ is smaller than $E_{dep}(t)$, and therefore it occurs in the opposite direction to that of $E_{appl}$ (Note that $E_{appl}$ and $E_{dep}$ always points in opposite directions during unipolar fatigue; in some special situations where $E_{appl}$ remains zero after poling backswitching is the main cause for polarization retention loss in ferroelectric thin films.[34]). Similarly, "switching" during unipolar fatigue is defined as a polarization switching process driven by the residual applied/fatigue field $[E_{appl}(t)- E_{dep}(t)]$ when $E_{appl}(t)$ is larger than $E_{dep}(t)$, and therefore it occurs in the same direction as that of $E_{appl}$. The maximum $E_{dep}$ in a poled short-circuited ordinary Pt/PZT/Pt thin-film capacitor with interfacial layers has been estimated to be ~85 kV/cm (it depends on film parameters),[34] equivalent to ~2.5 V for the film investigated in this work. Therefore, it becomes obvious that when $E_{appl}(t)>E_{dep}(t)$ switching process dominates, and when $E_{appl}(t)<E_{dep}(t)$ backswitching process takes over (see Fig 5). The repetitive switching-and-backswitching process eventually leads to the noteworthy unipolar fatigue rate during monopolar cycling observed in this work (Fig 6-8). Since the switching-and-backswitching process during unipolar fatigue is much less intensive in comparison with that during bipolar fatigue in terms of the total number of domains involved and the completeness of polarization reversal in the film, unipolar cycles are supposed to gives rise to little or less severe fatigue than bipolar ones. And that is exactly what we observed in this work. It should be noted that the conventional view that switching does not take place during unipolar electrical cycling may need to be revised. Also note that the explanations given above are consistent with the LPD-SICI model.[2, 20] This model also predicts that much thicker thin-film capacitors and bulk ferroelectrics should show little or no unipolar fatigue in comparison with thinner ones, because $E_{dep}(t) = \dfrac{d_i P(t)}{d \varepsilon_i \varepsilon_0}$ [where $d$ and $d_i$ are the thickness of the film and the interfacial layer, respectively. $\varepsilon_i$ is the interfacial dielectric constant. $P(t)$ is the time-dependent polarization][34] and it becomes negligible when $d \gg 100$ nm (therefore the repetitive backswitching-and-switching process disappears). This is consistent with the previous results on relatively thicker ferroelectric polymer films of 1 μm,[5] and the commercialized ceramics of 1 mm.[4] This model also indicates that Pt/SBT/Pt (SBT stands for $SrBi_2Ta_2O_9$) or $RuO_2$/PZT/$RuO_2$ thin-film capacitors with better interface properties[35] should show enhanced unipolar fatigue endurance in comparison with Pt/PZT/Pt structure.

The reason for the significant bipolar/unipolar fatigue at 3 V evaluated by PUND measurements (Fig 2 and 6) is probably due to a certain spread of the magnitude of the activation field $\alpha$ for switching.[22] The parts/regions with low $\alpha$ have more chances or higher backswitching probability to switch/backswitch than those with relatively higher $\alpha$ at low fields. So they phase decompose earlier. Therefore the 3 V fatigue state evaluated by PUND shown in Fig 2 and Fig 6 simply indicates that almost all (~90%) the parts/regions with low enough $\alpha$ that could be easily switched or backswitched at 3 V collapse after $10^9$ cycles, while those with higher $\alpha$ remain *almost* unaffected and they can readily be activated by higher voltages (e.g. 6-12 V used in hysteresis-loop measurements shown in Fig 3 and Fig 7). This confirms again our argument made above that we need to apply higher PUND voltages or use hysteresis-loop measurements at higher voltages when the fatigue voltage is about the coercive voltage or even lower.

Additionally, since $\int_0^{N \cdot T} V_{bipolar}(t) \cdot dt = \int_0^{N \cdot T} V_{unipolar}(t) \cdot dt$ for the same $N$ and $T$ (see Fig 1 and Fig 5; $N$ is the number of cycles defined above and $T$ is the period of one cycle; $T=1/f$. $f$ is fatigue frequency and is $10^5$ Hz in this work), it is expected that the total amount of charge passing through the capacitor would be about the same for both bipolar and unipolar fatigue after the same number of cycles subject to the same cycling voltage and frequency. So the different characteristics between bipolar and unipolar fatigue at the same voltage provides a strong evidence suggesting that bipolar/unipolar fatigue is induced by charge injection during switching, not that during post-switching stage, i.e., the stable or quasi-stable leakage current stage (for another evidence the interested readers is referred to the work on bipolar fatigue in antiferroelectrics[24]).

## D. Effective thickness of the phase-decomposed layer during bipolar/unipolar fatigue: an estimation based on dielectric measurements

Let us now evaluate the effective thickness of the degraded layer during bipolar/unipolar fatigue based on the dielectric data shown in Fig 4 and Fig 8, which could not be done previously using the micro-Raman data on fatigued PZT films.[13] To do this, we assume that the Pt/PZT/Pt structure could be described using a conventional in-series capacitor model, consisting of the interfacial degraded layer (which includes both the original passive layer between Pt and virgin PZT and the phase-decomposed layer induced by unipolar/bipolar fatigue; the thickness of the former is much less than that of the latter as we will see later on) and the ferroelectric bulk. Then we assume that the dielectric constant of the original passive layer (also called "dead layer" in some publications) is about the same as that of the phase-decomposed pyrochlorelike layer caused by fatigue as was justified previously (for simplicity both are labeled by $\varepsilon_i \sim 40$, one order of magnitude less than that of the bulk)[20]. [In a recent work, we argued that the passive layer between virgin PZT (or barium strontium titanate, BST) and metal electrodes like Pt and Au might have a defective pyrochlore/fluorite structure (possibly with a small portion of ferroelectric perovskite phase)[36]. By assuming $d_i<<d$ (or $\varepsilon_f>>\varepsilon_i$, $d_i$ is defined as the *effective thickness* of the interfacial degraded layer including the layer caused by fatigue. $\varepsilon_f$ is the dielectric constant of the bulk ferroelectric and is assumed to be unchanged during electrical cycling[7]), this model gives:[37]

$$\frac{1}{C} = \frac{d}{\varepsilon_f \varepsilon_0 S} + \frac{d_i}{\varepsilon_i \varepsilon_0 S} \qquad (1)$$

where $C$ is the measured capacitance, $S$ is capacitor area and $\varepsilon_0$ is vacuum permittivity. Defining $C \equiv \frac{\varepsilon \varepsilon_0 S}{d}$, we obtain:

$$\frac{d}{\varepsilon} = \frac{d}{\varepsilon_f} + \frac{d_i}{\varepsilon_i} \qquad (2)$$

where $\varepsilon$ is the measured dielectric constant of the capacitor. Eq (1) and (2) predict a linear relationship between $d/\varepsilon$ (or $1/C$) and $d$ with the slope of $1/\varepsilon_f$ and the y-axis intercept of $d_i/\varepsilon_i$. In a recent work on size effects in ferroelectric thin films,[36] the in-series capacitor model has been used to show that $\varepsilon_f$ ~450 and $d_i/\varepsilon_i$ ~0.18 for the Pt/PZT/Pt thin-film capacitors prepared by the similar fabrication method as those used in this work.

Using Eq (2) and the data in Fig 4 and Fig 8 (assuming $\varepsilon_f$ =450 and $\varepsilon_i$ =40 for simplicity as justified above), we could estimate $d_i(N)$ as a function of cycle number for both bipolar and unipolar fatigue at different fatigue voltages (see Fig 9). [Note that the effective thickness $d_i$ in Fig 9 should not be regarded as the *real thickness* of the interfacial degraded layer, because the phase-decomposed regions are around the nucleation sites, rather local and isolated at the beginning of fatigue process and may still not be perfectly planar even at the final stage of fatigue measurement subject to higher bipolar cycling voltages.[20]] Fig 9 provides another quantitative way to show that unipolar fatigue is indeed much less severe than bipolar fatigue subject to the same cycling voltage and the poorest fatigue endurance occurs for bipolar fatigue at 9 and 12 V, in good agreement with what we discussed in the previous sections.

Note that even after the most severe electrical cycling (e.g., after $10^9$ cycles of bipolar fatigue at 12 V) $d_i$ is ~39 nm only (Fig 9), still much less than the film thickness $d$~300 nm, indicating that most of the ferroelectric remains unaffected after fatigue. This confirms the view in our previous works that polarization fatigue in ferroelectric materials is an interface effect, not a bulk one.[2, 20]

## IV. Conclusions

In summary, we show that unipolar fatigue is noteworthy though still much less severe than bipolar fatigue subject to electrical cycling at the same voltage for the ferroelectric PZT thin-film capacitors deposited by sol-gel method. We attributed the reason for the evident unipolar fatigue in the thin-film samples observed in this work to polarization retention (or backswitching) induced by residual $E_{dep}$ between the monopolar pulses when $E_{dep}$ is larger than $E_{appl}$, according to the LPD-SICI model. So the conventional view that switching does not take place during monopolar electrical cycling may need to be changed. We also show that PUND data recorded using the pulses with the same voltage as those for repetitive unipolar/bipolar cycling are not reliable if the voltage is lower than $2V_c$ (equivalent to 6 V for our film). Hysteresis-loop measurements conducted at higher voltages (e.g. 12 V) or dielectric measurements are more reliable. Finally, using dielectric results we have estimated the effective thickness $d_i$ of the electrical-cycling-induced phase-decomposed (pyrochlorelike) interfacial layer during bipolar/unipolar fatigue, which has not been done so far. That fact that $d_i$ is ~39 nm, still much less than the film thickness $d$ ~300 nm even after the most severe bipolar fatigue strongly suggests that polarization fatigue in ferroelectrics is not a bulk effect, but an interface one.

**Acknowledgement**

X.J.L would like to thank the LKY PDF established under the Lee Kuan Yew Endowment Fund for support. The work is supported also by National University of Singapore and a MOE AcRF grant (R284-000-058-112). We thank Joe Evans and Scott Chapman for helpful discussion on the set-up of unipolar fatigue measurements.

**References:**

1    J. F. Scott, *Ferroelectric Memories* (Springer, New York, 2000).

2    X. J. Lou, Journal of Applied Physics **105**, 024101 (2009).

3    V. Chikarmane, C. Sudhama, J. Kim, J. Lee, A. Tasch, and S. Novak, Journal Of Vacuum Science & Technology A-Vacuum Surfaces And Films **10**, 1562 (1992).

4    C. Verdier, D. C. Lupascu, and J. Rodel, Journal Of The European Ceramic Society **23**, 1409 (2003).

5    G. D. Zhu, Z. G. Zeng, L. Zhang, and X. J. Yan, Applied Physics Letters **89**, 102905 (2006).

6    Radiant Technologies Inc., RT6000 Standardised ferroelectric test system operation manual.

7    J. J. Lee, C. L. Thio, and S. B. Desu, Journal of Applied Physics **78**, 5073 (1995).

8    T. Mihara, H. Watanabe, and C. A. P. de Araujo, Japanese Journal Of Applied Physics Part 1-Regular Papers Short Notes & Review Papers **33**, 5281 (1994).

9    A. Jiang, M. Dawber, J. F. Scott, C. Wang, P. Migliorato, and M. Gregg, Japanese Journal Of Applied Physics Part 1-Regular Papers Short Notes & Review Papers **42**, 6973 (2003).

10   Q. Y. Jiang, W. W. Cao, and L. E. Cross, Journal Of The American Ceramic Society **77**, 211 (1994).

11   D. Wang, Y. Fotinich, and G. P. Carman, Journal Of Applied Physics **83**, 5342 (1998).

12   C. Verdier, F. D. Morrison, D. C. Lupascu, and J. F. Scott, Journal Of Applied Physics **97**, 024107 (2005).

13      X. J. Lou, M. Zhang, S. A. T. Redfern, and J. F. Scott, Physical Review Letters **97**, 177601 (2006).

14      F. C. Kartawidjaja, Z. H. Zhou, and J. Wang, Journal of electroceramics **16**, 425 (2005).

15      M. Grossmann, D. Bolten, O. Lohse, U. Boettger, R. Waser, and S. Tiedke, Applied Physics Letters **77**, 1894 (2000).

16      T. Mihara, H. Watanabe, and C. A. P. de Araujo, Japanese Journal of Applied Physics Part 1-Regular Papers Short Notes & Review Papers **33**, 3996 (1994).

17      B. G. Chae, C. H. Park, Y. S. Yang, and M. S. Jang, Applied Physics Letters **75**, 2135 (1999).

18      K. Amanuma, T. Hase, and Y. Miyasaka, Japanese Journal Of Applied Physics **33**, 5211 (1994).

19      P. J. Schorn, D. Brauhaus, U. Bottger, R. Waser, G. Beitel, N. Nagel, and R. Bruchhaus, Journal Of Applied Physics **99**, 114104 (2006).

20      X. J. Lou, M. Zhang, S. A. T. Redfern, and J. F. Scott, Physical Review B **75**, 224104 (2007).

21      X. J. Lou, Journal of Physics-Condensed Matter **21**, 012207 (2009).

22      X. J. Lou, Journal of Applied Physics **105**, 094112 (2009).

23      Joe Evans (private communication).

24      X. J. Lou, Applied Physics Letters **94**, 072901 (2009).

25      W. J. Merz, Physical Review **95**, 690 (1954).

26      S. Hashimoto, H. Orihara, and Y. Ishibashi, Journal Of The Physical Society Of Japan **63**, 1601 (1994).

27      Y. W. So, D. J. Kim, T. W. Noh, J. G. Yoon, and T. K. Song, Applied Physics Letters **86**, 092905 (2005).

28      O. Lohse, M. Grossmann, U. Boettger, D. Bolten, and R. Waser, Journal Of Applied Physics **89**, 2332 (2001).

29      J. Y. Jo, H. S. Han, J. G. Yoon, T. K. Song, S. H. Kim, and T. W. Noh, Physical Review Letters **99**, 267602 (2007).

[30] Y. Ishibashi, in *Ferroelectric thin films: synthesis and basic properties*, edited by C. P. de Araujo, J. F. Scott and D. V. Taylor (Gordon and Breach Publishers, 1996), p. 135.

[31] Y. Ishibashi and Y. Takagi, Journal of the Physical Society of Japan **31**, 506 (1971).

[32] H. Orihara, S. Hashimoto, and Y. Ishibashi, Journal Of The Physical Society Of Japan **63**, 1031 (1994).

[33] A. K. Tagantsev, I. Stolichnov, N. Setter, J. S. Cross, and M. Tsukada, Physical Review B **66**, 214109 (2002).

[34] X. J. Lou, Journal of Applied Physics **105**, 094107 (2009).

[35] H. Z. Jin and J. Zhu, Journal Of Applied Physics **92**, 4594 (2002).

[36] X. J. Lou and J. Wang, submitted (2009).

[37] K. Amanuma, T. Mori, T. Hase, T. Sakuma, A. Ochi, and Y. Miyasaka, Japanese Journal Of Applied Physics Part 1-Regular Papers Short Notes & Review Papers **32**, 4150 (1993).

**Figure Captions:**

Fig 1 (color online) a schematic of bipolar fatigue signal profile

Fig 2 (color online) Bipolar fatigue evaluated by PUND measurements: (a) $P_{sw}(\equiv dP)$ as a function of cycle number $N$ for bipolar fatigue at 3V, 6V, 9V and 12V; (b) the corresponding normalized results.

Fig 3 (color online) Bipolar fatigue evaluated by hysteresis-loop measurements at 12 V after 1, $4\times10^5$, $1.9\times10^6$, $1.15\times10^7$, $1.02\times10^8$ and $1\times10^9$ cycles for bipolar fatigue at (a) 3 V, (b) 6 V, (c) 9 V and (d) 12 V; (e) $2P_r(N)/2P_r(0)$ as a function of cycle number $N$ (the normalized results).

Fig 4 (color online) Bipolar fatigue evaluated by dielectric measurements (with ac amplitude of 0.05 V and a frequency range from $10^1$ to $10^6$ Hz) after 1, $4\times10^5$, $1.9\times10^6$, $1.15\times10^7$, $1.02\times10^8$ and $1\times10^9$ cycles for bipolar fatigue at (a)(b) 3 V, (c)(d) 6 V, (e)(f) 9 V and (g)(h) 12 V; (a)(c)(e)(g) show the dielectric constant and tan loss data as a function of measurement frequency; (b)(d)(f)(h) display the dielectric constant and tan loss data measured at $10^2$ and $10^5$ Hz as a function of cycle number $N$. (i) normalized dielectric constant measured at $10^5$ Hz as a function of cycle number $N$ for bipolar fatigue at 3 V, 6 V, 9 V and 12 V.

Fig 5 (color online) a schematic of unipolar fatigue signal profile

Fig 6 (color online) Unipolar fatigue evaluated by PUND measurements: (a) $P_{sw}$ as a function of cycle number $N$ for unipolar fatigue at 3V, 6V, 9V and 12V; (b) the corresponding normalized results.

Fig 7 (color online) Unipolar fatigue evaluated by hysteresis-loop measurements at 12 V after 1, $4\times10^5$, $1.9\times10^6$, $1.15\times10^7$, $1.02\times10^8$ and $1\times10^9$ cycles for unipolar fatigue at (a) 3 V, (b) 6 V, (c) 9 V and (d) 12 V; (e) $2P_r(N)/2P_r(0)$ as a function of cycle number $N$ (the normalized results).

Fig 8 (color online) Unipolar fatigue evaluated by dielectric measurements (with ac amplitude of 0.05 V and a frequency range from $10^1$ to $10^6$ Hz) after 1, $4\times10^5$, $1.9\times10^6$, $1.15\times10^7$, $1.02\times10^8$ and $1\times10^9$ cycles for unipolar fatigue at (a)(b) 3 V, (c)(d) 6 V, (e)(f) 9 V and (g)(h) 12 V; (a)(c)(e)(g) show the dielectric constant and tan loss data as a function of measurement frequency; (b)(d)(f)(h) display the dielectric constant and tan loss data measured at $10^2$ and $10^5$ Hz as a function of cycle number *N*. (i) normalized dielectric constant measured at $10^5$ Hz as a function of cycle number *N* for unipolar fatigue at 3 V, 6 V, 9 V and 12 V.

Fig 9 (color online) $d_i$ (nm), the effective thickness of the interfacial passive/degraded layer, as a function of cycle number *N* during bipolar/unipolar fatigue at various voltages

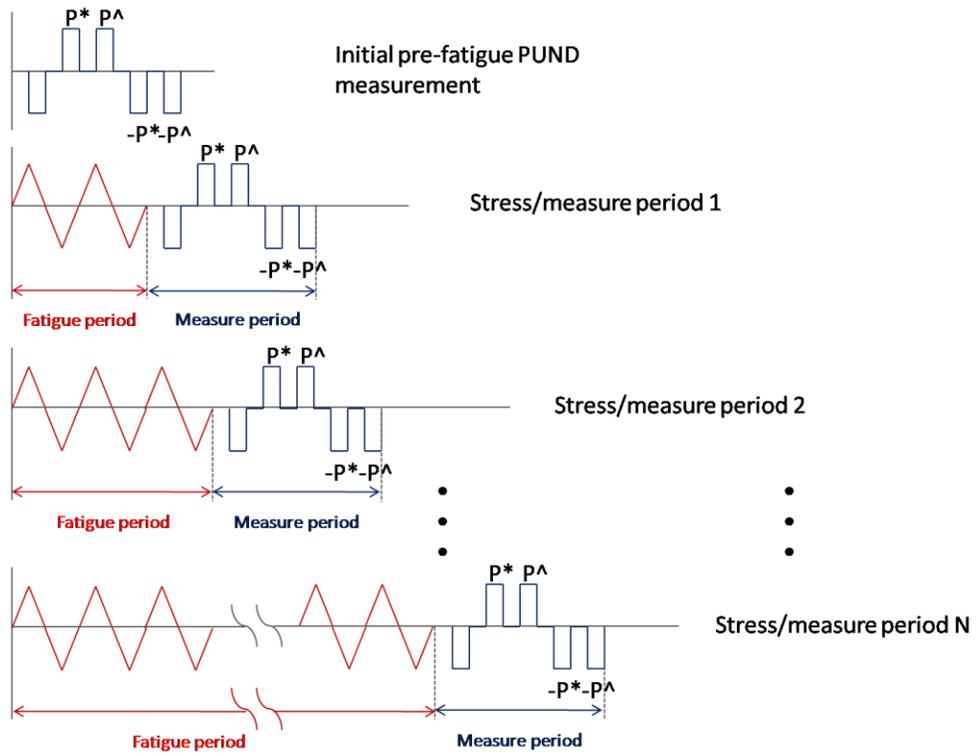

Fig 1

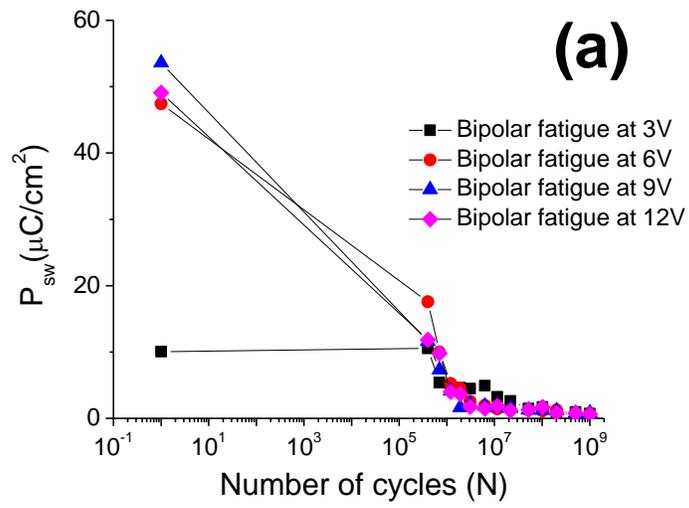

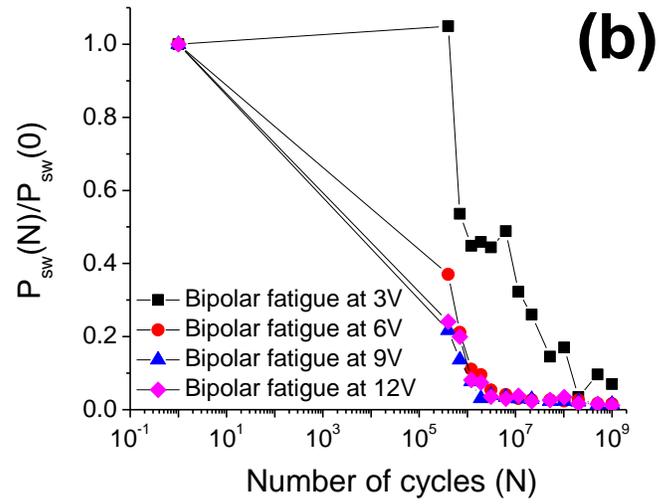

Fig 2

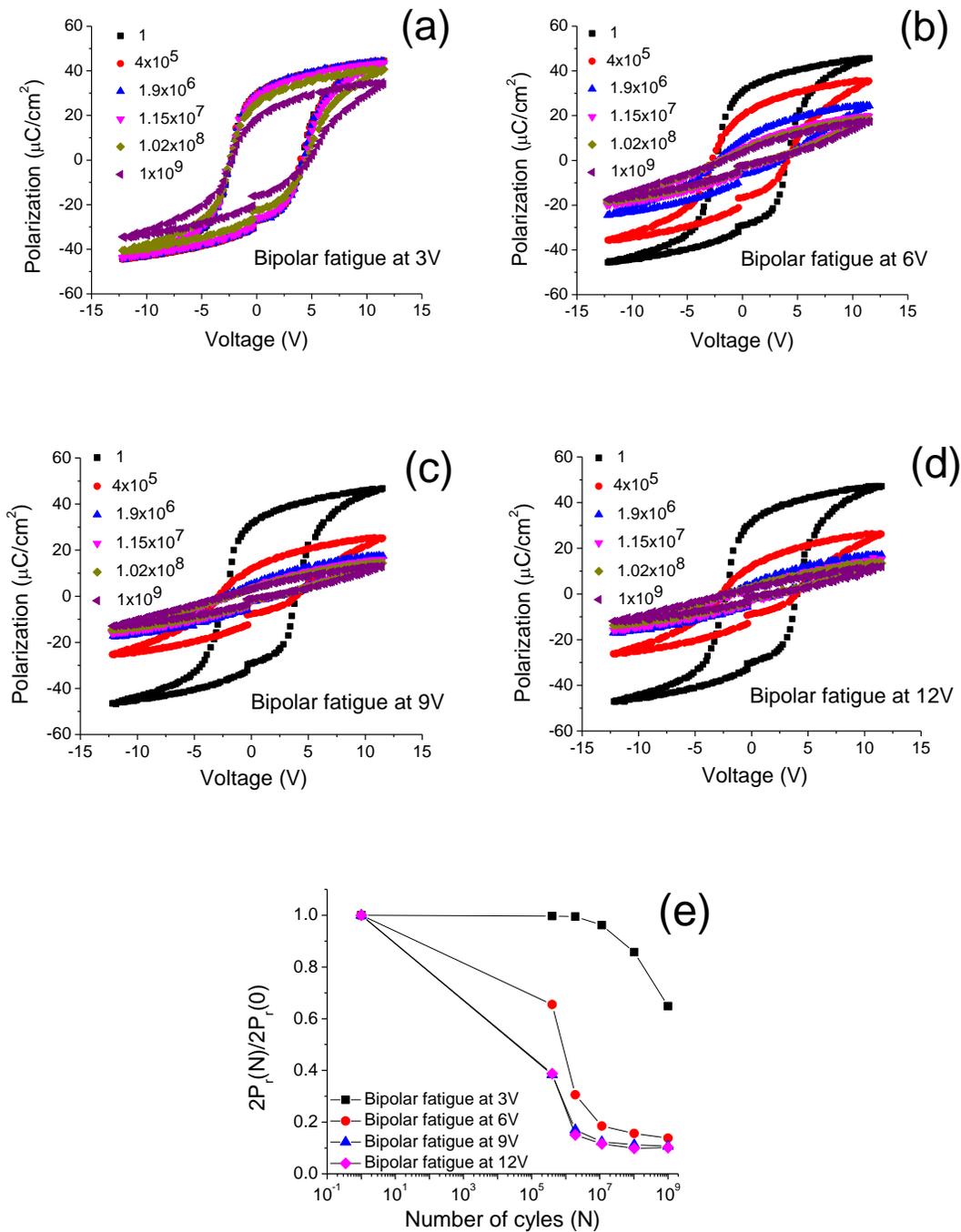

Fig 3

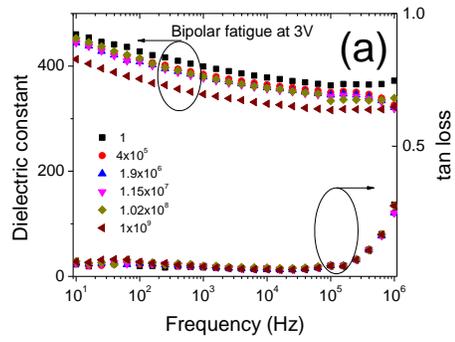
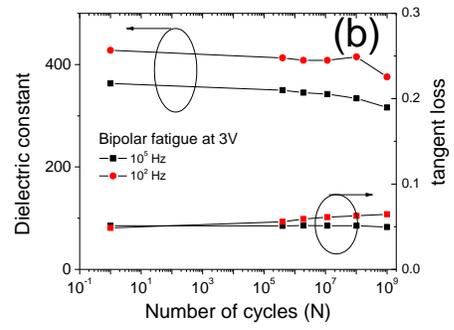
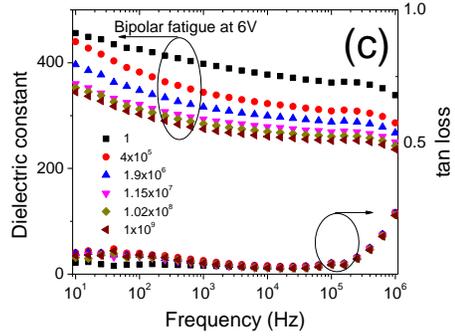
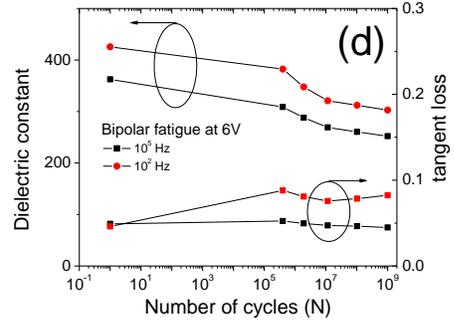
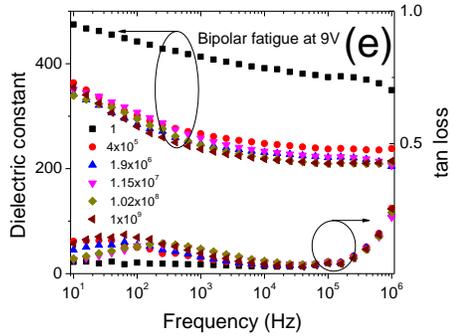
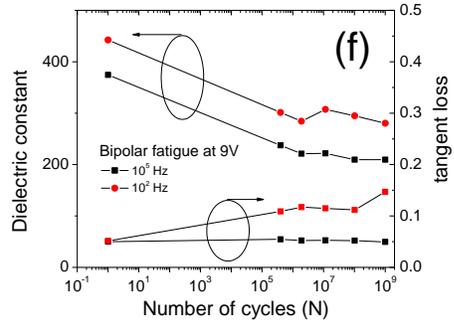
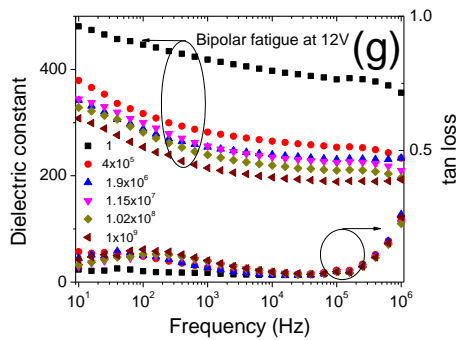
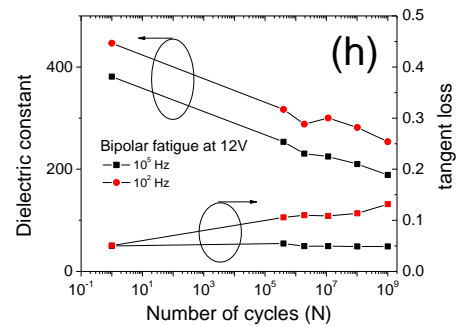

Fig 4 (to be continued)

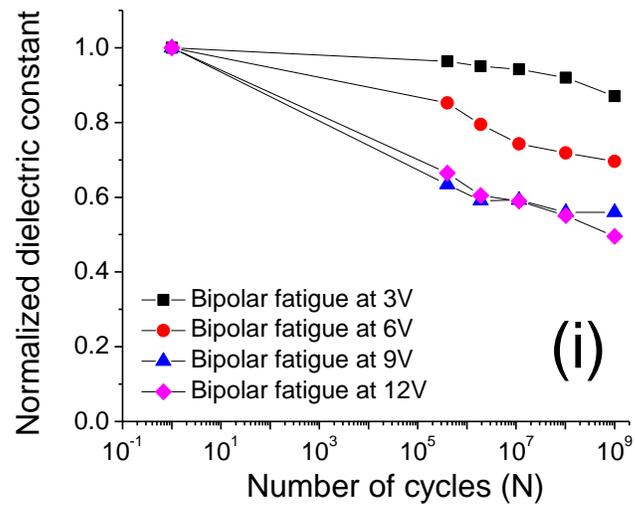

Fig 4

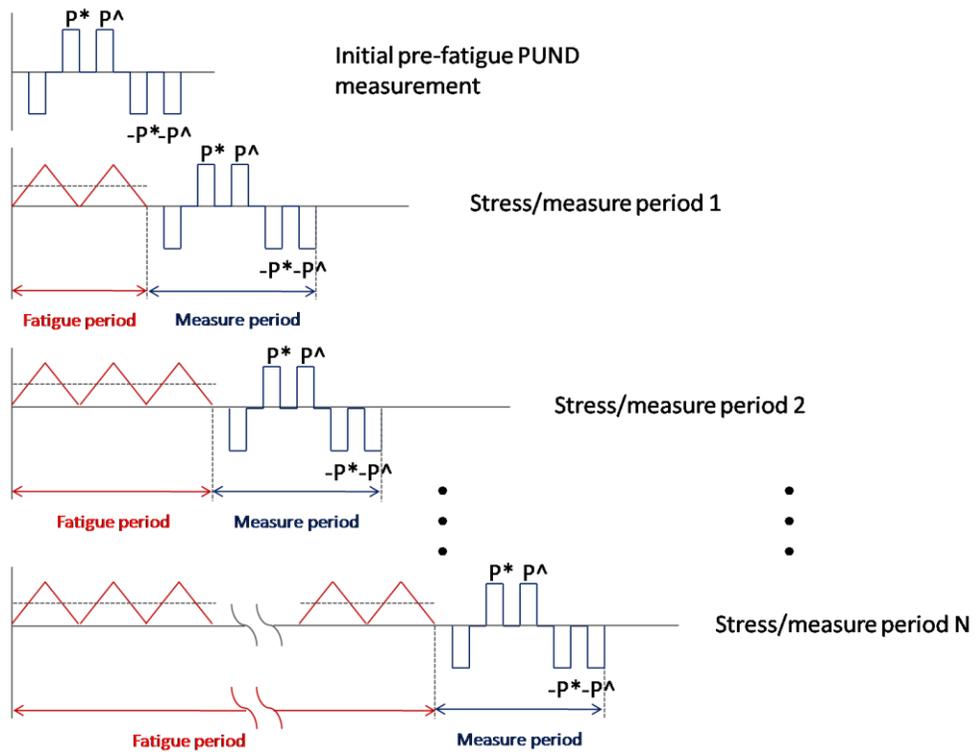

Fig 5

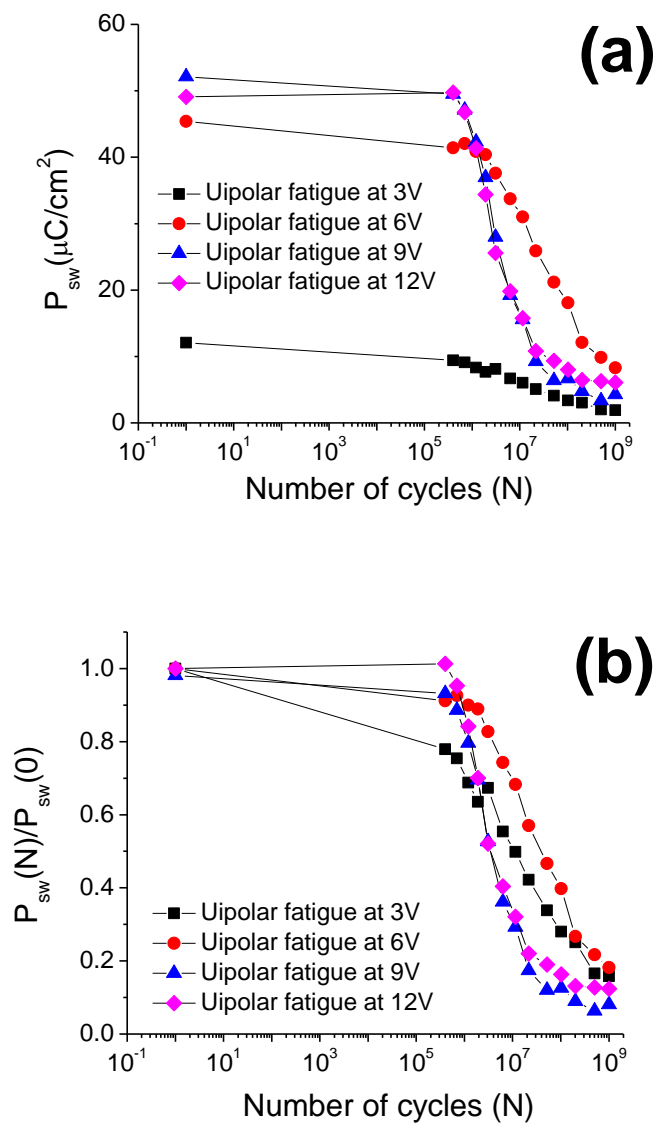

Fig 6

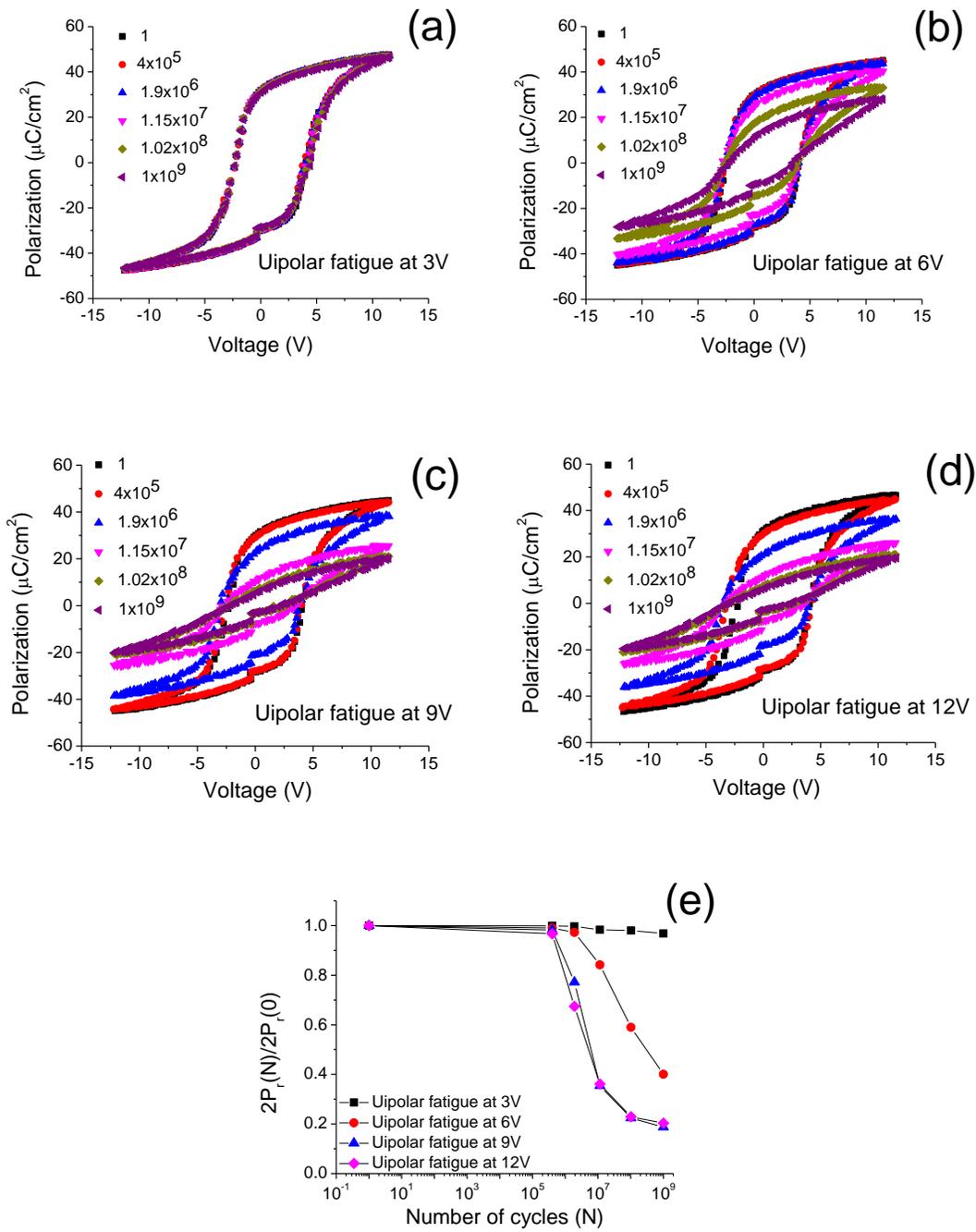

Fig 7

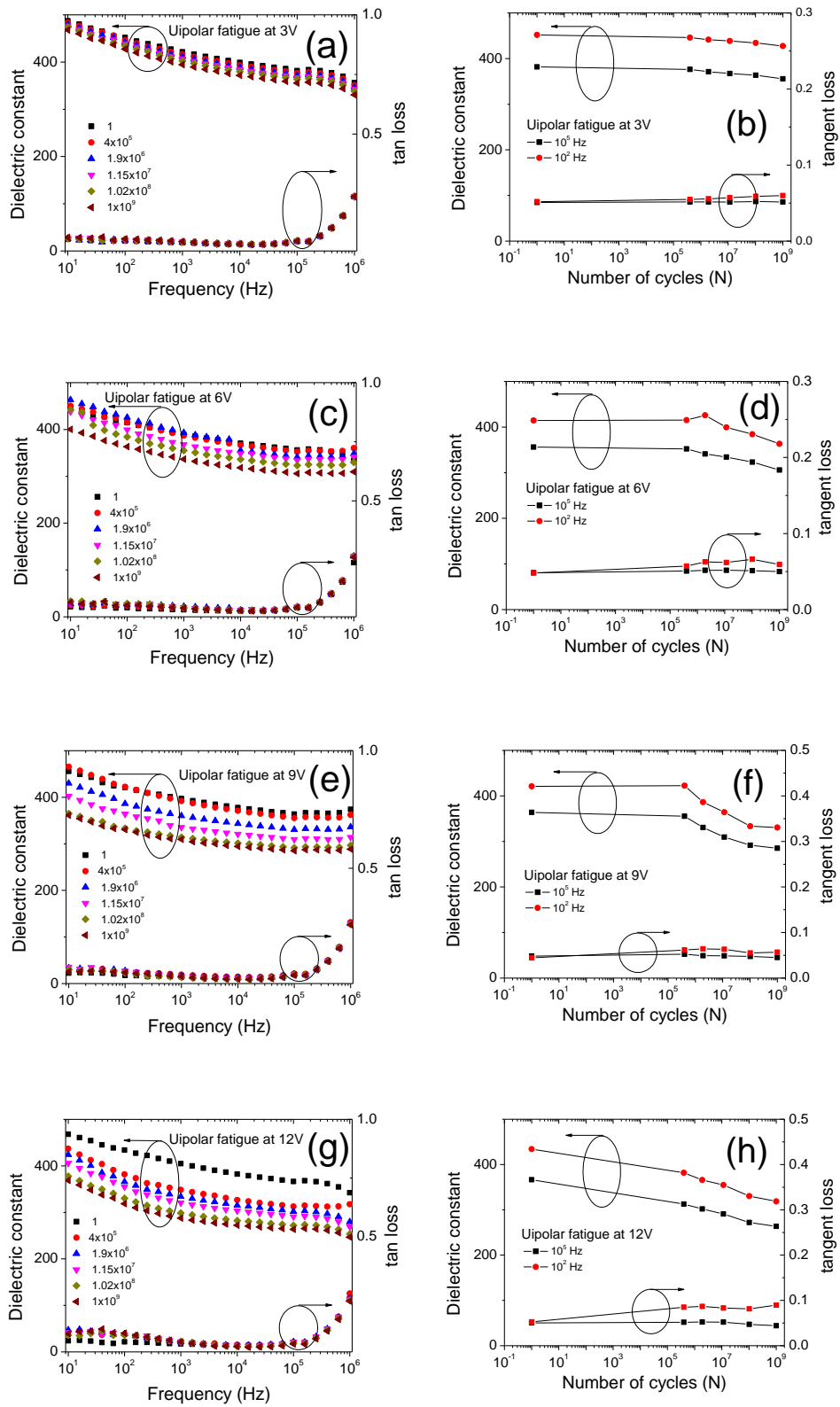

Fig 8 (to be continued)

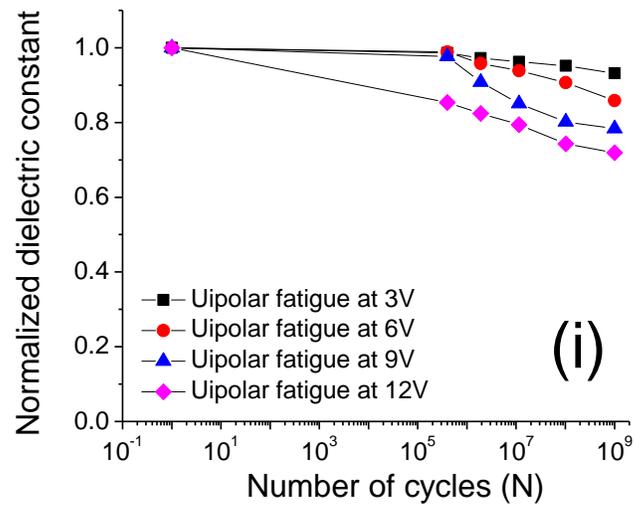

Fig 8

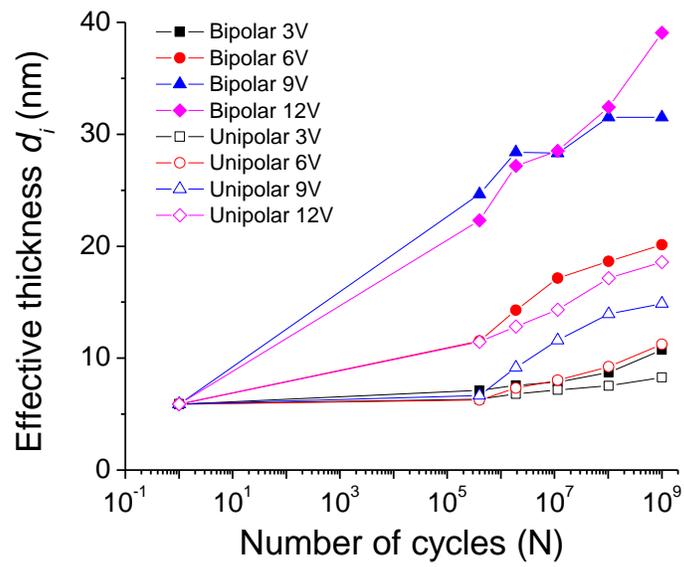

Fig 9